\documentclass[sigchi]{acmart}

\AtBeginDocument{%
  \providecommand\BibTeX{{%
    \normalfont B\kern-0.5em{\scshape i\kern-0.25em b}\kern-0.8em\TeX}}}

\setcopyright{acmcopyright}
\copyrightyear{}
\acmYear{}
\acmDOI{}

\acmConference{}
\acmBooktitle{}
\acmPrice{}
\acmISBN{}

\begin{document}

\title{Checking Fact Worthiness using Sentence Embeddings}

\author{Sidharth Singla}
\email{s6singla@uwaterloo.ca}
\affiliation{%
  \institution{University of Waterloo}
  \city{Waterloo}
  \state{Ontario}
}


\begin{abstract}
Checking and confirming factual information in texts and speeches is vital to determine the veracity and correctness of the factual statements. This work was previously done by journalists and other manual means but it is a time-consuming task. With the advancements in Information Retrieval and NLP, research in the area of Fact-checking is getting attention for automating it. CLEF-2018 and 2019 organised tasks related to Fact-checking and invited participants. This project focuses on CLEF-2019 Task-1 Check-Worthiness and experiments using the latest Sentence-BERT pre-trained embeddings, topic Modeling and sentiment score are performed. Evaluation metrics such as MAP, Mean Reciprocal Rank, Mean R-Precision and Mean Precision@N present the improvement in the results using the techniques. 
\end{abstract}

\keywords{Fact checking; Check worthiness; Sentence-BERT; Speech}

\maketitle

\section{Introduction}
Fact-checking is a process that seeks to investigate an issue in order to verify the facts, according to the Oxford dictionary. The spread of misinformation is inherently human. Important facts can be turned and twisted both knowingly and unknowingly and can lead to the spread of misinformation in the masses.
Fake News detection is thus becoming a hot topic of research. Thus fact-checking and fake news detection is very much related to each other.\\
Political leaders, for their campaigns and rallies, tend to present twisted facts with incorrect and disguised figures. This particularly caught public attention during the 2016 US Presidential Campaign which was influenced by fake news and false claims. It becomes essential to check if the facts being presented are correct or not. Also, it is an important task to determine the sentences or phrases out of text and speeches that should be factually-checked. The manual process is cumbersome as one has to go through long text and speeches and check each of the sentences if it needs to be checked. Also, after extracting the phrases to be looked up for fact-checking it is required to check them in order of the relevance of the phrase. Few phrases will be more relevant to check for and should first become a priority for validation. Doing this manually is very time-consuming and there is a lot of possibility of human error.\\
With the social media getting stronger and used, a statement, an interview, a press release, a blog note, or a tweet can spread almost instantaneously. With the mass of information and little time for
double-checking claims against the facts, it becomes essential to have a quicker process and system for validation. A number of organisations such as FactCheck and Snopes, among many others have started with fact-checking initiatives but it has proved to be a very demanding manual effort and only small number of claims could be fact-checked.\\
This motivates researchers and engineers to research and build automatic intelligent systems that are able to detect the statements that are important to be fact-checked, rank them in order of priority and then check if the fact statement is correct or twisted.\\

CheckThat! Lab at CLEF-2019\cite{clef_19} aimed to address these problems and the participants were asked to build systems that help in automating the systems. It states that a typical fact-checking pipeline includes the following steps.
\begin{enumerate}
    \item Identification of check-worthy text fragments.
    \item Retrieval of documents that might be useful for fact-checking the claim from various sources with the extraction of supporting evidence.
    \item Determination if the claim is true or not by comparing a claim against the retrieved evidence.
\end{enumerate}

The \textbf{Task 1}\cite{clef_2019_task1} is a ranking task. Given a political debate or a transcribed speech, segmented into sentences with speakers annotated, the task is to identify which sentence should be prioritized for fact-checking. The systems are required to produce a score per sentence, according to which the ranking will be performed and the task runs in English.\\

Eleven teams participated in Task 1. One of the participant team named \textbf{TOBB ETU}\cite{tobb_etu} used linguistic features such as named entities, topics
extracted with IBM Watson’s NLP tools, PoS tags, bigram counts and trained a multiple additive regression tree(MART).\\
This project improves their results by using Sentence BERT\cite{sbert} embeddings for the sentences and using sentiment scores and topic modeling scores of a sentence as features. Data Augmentation was also used to create more sentence rows for the dataset. All the experiments and results are presented.

\section{Related Work}

\subsection{Fact-checking systems}
The very first work done to check-worthiness was the ClaimBuster \cite{claimbuster} system. TF-IDF word representations were used with the features such as sentiment, part-of-speech (POS) tags and named entities. SVM trained classifier model was trained on data manually annotated by students, professors,
and journalists, where each sentence was annotated as non-factual, unimportant factual, or check-worthy factual. The data consisted of transcripts of 30 historical US election debates covering the period from 1960 until 2012 for a total of 28,029 transcribed sentences.It was later evaluated against CNN and PolitiFact. Patwari et al.\cite{patwari} presented a system called TATHYA based on the similar features, and also included contextual features sentences immediately preceding and succeeding the one being assessed, as well as certain hand-crafted POS patterns. Gencheva et al.\cite{gencheva} extended ClaimBuster's feature set by including more contextual features, such as the sentence’s position in the debate text, and whether the debate opponent is mentioned. Konstantinovskiy et al.\cite{konstantinovskiy} in their work used neural networks for checking worthiness. They used InferSent\cite{infersent} for the universal neural sentence representation. A logistic regression classifier was then trained. Thorne et al.\cite{thorne} presented automatic fact-checking as a multi-step process that includes: identifying check-worthy statements; generating questions to be asked about the statements; retrieving relevant information to create a knowledge base; inferring the veracity of the statements using text analysis or external sources. Vasileva et al.\cite{vasileva} proposed a multi-task learning neural network that learns from nine fact-checking organizations simultaneously and
predicts if a sentence will be selected for fact-checking by each of these organizations.

\subsection{CLEF2019 Task1 Submissions}
In this subsection, approaches used by the participants for the CLEF-19 Task1 primary subtask is discussed. Team \textbf{Copenhagen}\cite{copenhagen} achieved the best performance. The system learned dual token embeddings: domain-specific word embeddings and syntactic dependencies, and used them in an LSTM network. They pre-trained this network with previous Trump and Clinton debates, and then supervised it weakly with the ClaimBuster system. In their primary submission, they used a contrastive ranking loss.
Team \textbf{TheEarthIsFlat}\cite{earth_is_flat} trained a feed-forward neural network with two hidden layers, which takes as input Standard Universal Sentence Encoder(SUSE) embeddings for the current sentence as well as for the two previous
sentences as a context. Team \textbf{IPIPAN}\cite{ipipan} extracted features such as bag-of-words n-grams, word2vec vector representations, named entity types, part-of-speech tags, sentiment scores, and features from statistical analysis of the sentences. These features were then used in an L1-regularized logistic regression to predict the check-worthiness of the sentences. Team \textbf{Terrier}\cite{terrier} represented the sentences using bag-of-words and named entities and used co-reference resolution to substitute the pronouns by the referring entity/person name. They also computed entity similarity and entity relatedness and used an SVM classifier. Team \textbf{UAICS}\cite{uaics} used a Naive Bayes classifier with bag-of-words features. Team \textbf{Factify} used pre-trained ULMFiT model and fine-tuned it on the training set. They over-sampled the minority class by replacing words randomly based on word2vec similarity and also used data augmentation based on back-translation, where each sentence was translated to French, Arabic and Japanese and then back to English. Team \textbf{JUNLP}\cite{junlp} extracted features such as syntactic n-grams, sentiment polarity, text subjectivity, and LIX readability score, and used them to train a logistic regression classifier with high recall. Then, they trained an LSTM model having GloVe word representations and part-of-speech tags.
The sentence representations from the LSTM model were concatenated with the extracted features and used for prediction by a fully connected layer, which had high precision. Finally, average posterior probabilities from both models were used as the final check-worthiness score. Team \textbf{nlpir01} extracted TF-IDF word vectors, TF-IDF PoS vectors, word, character, and PoS tag counts. These features were then used in a multilayer perceptron regressor with two hidden layers. Team \textbf{IIT/ISM Dhanbad} trained an LSTM network and fed the network with word2vec embeddings and features extracted from constituency parse trees as well as features based on named entities and sentiment analysis. Team \textbf{´e proibido cochilar} trained an SVM model on BoW representations of the sentences, after performing co-reference resolution and removing all digits. They further used an additional corpus of labelled claims extracted from fact-checking websites to have a more balanced training corpus and potentially better generalizations. The method used by team \textbf{TOBB ETU}\cite{tobb_etu} is discussed in the next subsection.

\subsection{Approach used by TOBB ETU}
The team used a hybrid approach in which claims are ranked using a supervised method and then reranked based on hand-crafted rules. MART was used as learning-to-rank algorithm to rank the claims. The features included topical category of claims, named entities, partof-speech tags, bigrams, and speakers of the claims. Rules were developed to detect the statements that are not likely to be a claim, and the statements were put at the very end of the ranked lists. The primary model ranked 9th based on
MAP, and 6th based on R-P, P@5, and P@20 metrics in the official evaluation of primary submissions. \textbf{Named Entities} were used as claims about people and institutions are likely to be check-worthy. They detected whether a statement is about a person or an institution, and existence of numerical values, location, and date information. Named entities were identified using Stanford Named Entity Tagger2 which tags entities such as person, location, organization, money, date, time and percentage. \textbf{Part-of-speech Tags} of words is helpful in detecting the amount of information in a sentence. For instance, a sentence having a noun tag is expected to have information to be fact-checked.
POS tags are detected using Stanford POS toolkit\cite{stanford_pos}. \textbf{Topical Category} is an effective indicator for check-worthy claims. For instance, a claim about celebrities can be considered less check-worthy than a claim about economics or wars. The topical categories such as finance, law, government, and politics, where each topic could be branched up to two levels of subtopics of statements was detected using IBM-Watson’s Natural Language Understanding Tool. A limited but powerful set of \textbf{bigrams} were used instead of using a large. Bigrams appearing at least N times only in check-worthy or only in not check-worthy claims in the training set were considered. N was set to 50.\\
The contribution of each of the feature was shown by the authors by performing ablation studies.

\subsection{Sentence-BERT}
Sentence-BERT(SBERT)\cite{sbert} is a modification of the pretrained BERT network and uses siamese and triplet network structures to derive semantically meaningful sentence embeddings that can be compared using cosine-similarity. The effort for finding the most similar pair from 65 hours with BERT / RoBERTa is reduced to about 5 seconds with SBERT, still maintaining the accuracy from BERT. SBERT and SRoBERTa  is evaluated on common STS tasks and transfer learning tasks, where it outperforms other state-of-the-art sentence embeddings methods. SBERT is fine-tuned on NLI data, which creates sentence embeddings that significantly outperform other state-of-the-art sentence embedding
methods like InferSent\cite{infersent} and
Universal Sentence Encoder on
seven Semantic Textual Similarity (STS) tasks.

\section{Dataset}
The dataset for Task 1 is an extension of the Check That-18 dataset\cite{check_that_18} and the combined training and test English part of 2018 dataset is the training data for 2019 Task. For the new test set, labelled data was produced from three press conferences, six public speeches, six debates, and one post. The annotations for the new instances were derived from the publicly available analysis carried out by factcheck.org. The annotation is at the sentence level. Therefore, if only part of a sentence was fact-checked, the entire sentence was annotated as a positive instance. If a claim spans more than one sentence, all the sentences are annotated as positive. The participating systems were allowed to use external datasets with fact-checking related annotations as well as to extract information from the Web, from social media, etc.

\begin{figure}[h]
  \centering
  \includegraphics[width=\linewidth, height=275px]{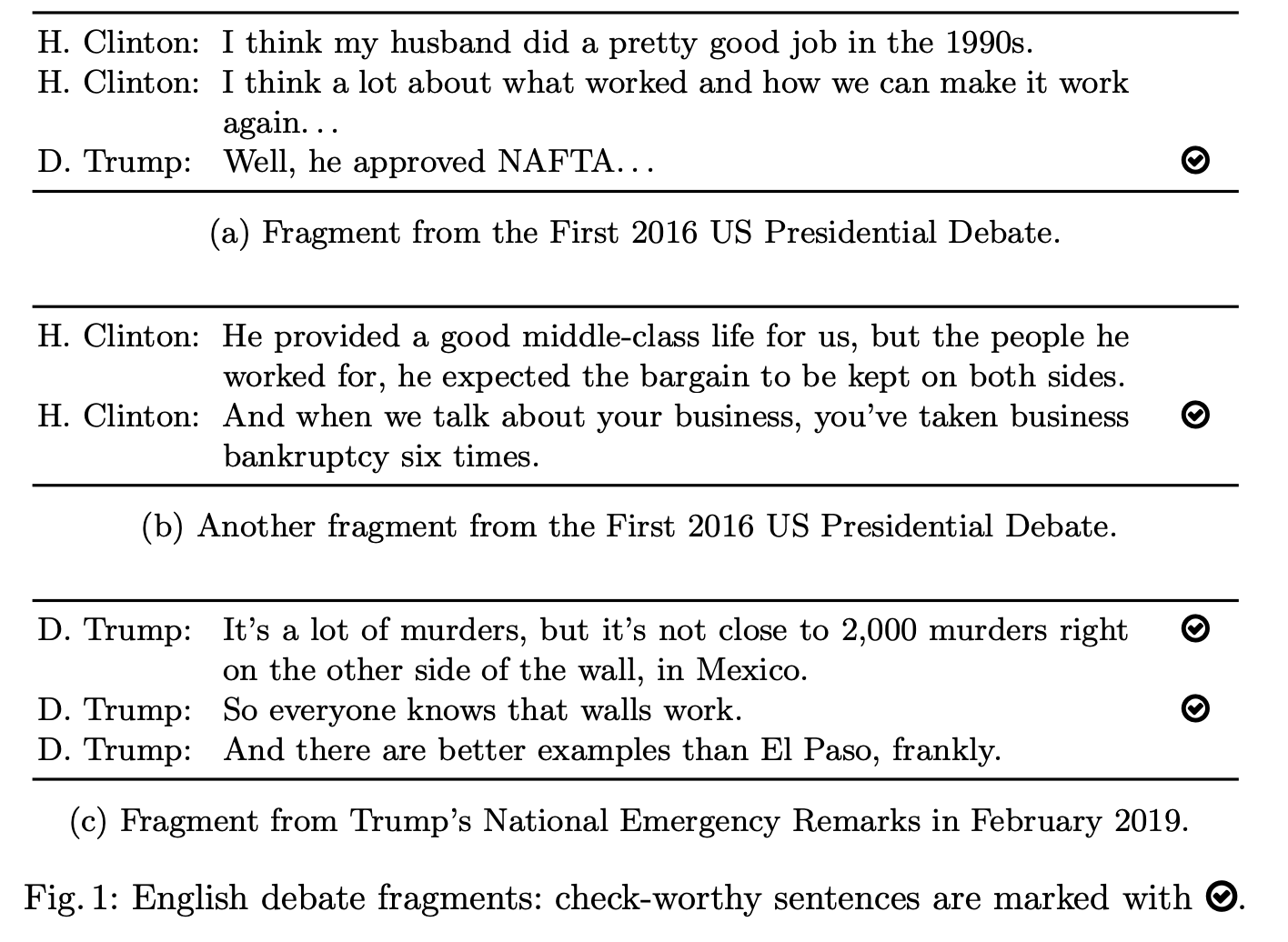}
  \caption{Dataset Training example}
\end{figure}

\section{Approach}
\begin{table*}[t]
  \caption{ System Performance after using SBERT Embeddings, Sentiment Features and Topic Modeling Features and combining with TOBB ETU's system.}
  \label{tab:results}
  \begin{tabular}{cccccccccc}
    \toprule
     & MAP & RR & R-P & P@1 & P@3 & P@5 & P@10 & P@20 & P@50\\
    \midrule
    TOBB ETU & .0884 &	.2028 &	.1150 &	.0000	& .0952 &	.1429 &	.1286 &	.1357 &	.0829\\
    
    SF only & .0832 & .3017 & .0873 & .1429 & .1905 & .1429 & .1286 & .1000 & .0800\\
    
    SF + TOBB ETU & .0885 & .1992 & .1218 & .0000 & .1429 & .1143 & .1286 & .1500 & .0829\\
    
    SBERT only & .1243 & .2207 & .1522 & .1429 & .0952 & .0857 & .1571 & .1643 & .1200\\
    
    SBERT + TOBB ETU & .1287 & .2318 & .1577 & .1429 & .1429 & .1714 & .1429 & .1786 & .1171\\
    
    TMF only & .1151 & .3487 & .0979 & .2857 & .1905 & .1714 & .1429 & .1071 & .0800\\
    
    TMF + TOBB ETU & .1063 & .1930 & .1327 & .0000 & .0952 & .2000 & .1714 & .1429 & .0886\\

    SBERT + TMF + SF +  TOBB ETU & .1396 & .2780 & .1348 & .1429 & .1905 & .1714 & .1571 & .1643 & .1171\\
  \bottomrule
    \end{tabular}
\end{table*}
The TOBB ETU's performance is improved in this project by using additional features. The training and testing algorithm of the original system remains the same and only new features are added for the performance. The original system uses the MART algorithm with 50 trees and 2 leaves and the same is used in this approach. One huge advantage of the approach is that training time is very less and the results are also obtained quickly as there is no neural model training involved. The system doesn't require any GPU. Following features were implemented.

\subsection{Sentence BERT Embeddings} 
Each sentence is converted to the Sentence BERT (SBERT) Embeddings for both training and test files.
The model 'bert-base-nli-mean-tokens' is used for generating the embeddings. The model generates 768 sized vector embeddings for a sentence. These embeddings are the best to be used as representations as they capture the semantic meaning of a sentence.

\subsection{Sentiment Score}
Sentiment Analysis is the process of ‘computationally’ determining whether a piece of writing is positive, negative or neutral.
Sentiment score for each sentence is calculated using NLTK's VADER (Valence Aware Dictionary and sEntiment Reasoner) SentimentIntensityAnalyzer module for both the training and testing files. The analyzer computes negative, neutral and positive sentiment scores for each of the sentences and is added as features. 

\subsection{Topic Modeling}
\begin{table*}[t]
        \caption{ CLEF-19 Task 1 Top 3 submissions scores compared to the presented model}
        \label{tab:clef_results}
        \begin{tabular}{cccccccccc}
        \toprule
        & MAP & RR & R-P & P@1 & P@3 & P@5 & P@10 & P@20 & P@50\\
        \midrule
        Copenhagen & .1660 & .4176 & .1387 & .2857 &	.2381 &	.2571 &	.2286 &	.1571 &	.1229\\
    
        TheEarthIsFlat & .1597 & .19531 & .2052 &	.0000 &	.0952 &	.2286 &	.2143 &	.1857 &	.1457\\
    
        Presented System & .1396 & .2780 & .1348 & .1429 & .1905 & .1714 & .1571 & .1643 & .1171\\
        
        IPIPAN	& .1332	& .2864	& .1481	& .1429 &	.0952 &	.1429 &	.1714 &	.1500 &	.1171\\
    
         TOBB ETU & .0884 &	.2028 &	.1150 &	.0000	& .0952 &	.1429 &	.1286 &	.1357 &	.0829\\
        \bottomrule
        \end{tabular}
    \end{table*}
Topic Modeling is done with the sentences having label '1' with the LDA algorithm. 

Latent Dirichlet allocation (LDA): It is an unsupervised learning approach for Topic Modeling and a generative probabilistic model of a corpus. Documents are represented as random mixtures over latent topics, where each topic is characterized by a distribution over words.\\
Stopwords and punctuations are removed from the sentences and the retrieved phrases are converted to Doc2Bow (Document to Bag of Words) model using Gensim’s dictionary and LDA is performed after it. 
The number of topics extracted is 40 and top 5 words with their scores are considered.
Most of the words in the topics contained Nouns.
Feature vectors are created with the scores of the words present in the topics. The scores of the words not in the topics is considered as 0. From the results, it can be seen that it has improved the performance considerably.

\subsection{Data Augmentation}
Data Augmentation was performed by replacing Noun and Adjective words in a sentence with the corresponding most similar word using word2vec similarity. New generated sentences were added to the training dataset. But the results did not improve much and hence, are not reported.

\section{Evaluation and Results}
Following metrics are considered for the task evaluation. Each metric is described.

\subsection{Mean Average Precision (MAP)}
It is the official metric used in the competition for ranking. It is the mean of the average precision scores for each query.
\begin{equation}
    MAP = \sum_{q=1}^{Q} AveP(q)/Q
\end{equation}

\subsection{Reciprocal Rank}
Reciprocal rank\cite{reciprocal_rank} of a query response is the multiplicative inverse of the rank of the first correct answer. Mean reciprocal rank is the average of the reciprocal ranks of results for a sample of queries Q.
\begin{equation}
    MRR = 1/\left | Q \right |\sum_{i=1}^{\left | Q \right |}1/rank\textsubscript{i}
\end{equation}

\subsection{R-Precision}
R-Precision\cite{r_precision} is Precision at R, where R is the number of relevant line-numbers for the evaluated set. R-precision requires knowing all documents that are relevant to a query. If there are r relevant documents among the top-R retrieved documents,
\begin{equation}
    R-precision = r/R.
\end{equation}

\subsection{Precision@N}
Precision@N\cite{precision_n} is precision estimated for the first N line-numbers in the provided ranked list. It is the proportion of the top-n documents that are relevant. If r relevant documents have been retrieved at rank N,
\begin{equation}
    Precision@N = r/N
\end{equation}

Table \ref{tab:results} shows the results after adding the features on top of the TOBB ETU's implementation.

In the table:
\begin{itemize}
    \item SF refers to Semantic Features, the features extracted by performing sentiment analysis on a sentence.\\
    
    \item TMF refers to the Topic Modeling features, extracted by using Topic Modeling scores for each sentence.\\
    
    \item TOBB ETU refers to the original system built by the TOBB ETU's team.
\end{itemize}

From the results, it can be seen that the approach and features used improve the scores of all the evaluation metrics. The ablation studies and experiments also show the contribution and improvement provided by each of the features. The following analysis can be done from the table:

\begin{itemize}
    \item It can be seen that building a system using only Sentence BERT embeddings (Row 4) provide very good results. Combining the embeddings and using TOBB ETU's system (Row 5) improves the performance. The performance for the MAP score has improved by \textbf{45.58\%}.\\
    
    \item The sentiment features have not contributed much to the results. Comparing Rows 1 and 3, it can be seen that the metric results for TOBB ETU's system and TOBB Etu's system with the Sentiment Features are very much similar. Row 2 presents the scores when only Sentiment Features are used in the system.\\

    \item Topic Modeling Features also contribute greatly to the results. Row 6 shows the scores when only Topic Modeling features are used. Combining the features with TOBB ETU's system (Row 7) improves the performance in MAP score by \textbf{20.24\%}.\\
    
    \item Combining all the features: SBERT embeddings, Topic Modeling Features and Semantic Features with the TOBB ETU's system improves the performance of original TOBB ETU's system by \textbf{57.91\%} in MAP score, \textbf{17.21\%} in R-P score and \textbf{41.25\%} in Precision@50.\\
    
    Compared to CLEF-2019 competition TASK-1 submissions (Table \ref{tab:clef_results}), the MAP score of the presented system is the third best compared to the ninth spot of the TOBB ETU's system. The top 2 submissions are neural-based models. The team Copenhagen have used LSTM recurrent network as the model and the team TheEarthisFlat has used Feedforward network as the model. Comparatively, the presented approach doesn't use any neural model in training and still achieves very good performance and the results are comparable to the other two.
\end{itemize}

\section{Conclusion}
Here, a system is presented for fact-checking. CLEF-2019 CheckThat Lab's Task - 1 Check Worthiness! is the focus of the project that checks if a statement is worth fact-checking and ranks the statements according to which they should be prioritized. The system presented is built on top of the submission by TOBB ETU's team in the challenge and improves the metrics scores of their system by 57.91\% in Mean Average Precision Score by combining their implementation with features such as Sentence BERT embeddings, Topic Modeling Feature Scores and Semantic Feature scores. The contribution and performance of each of the feature is also presented as part of the ablation studies. Data Augmentation was also experimented but it didn't improve the results. Also, another advantage of the presented approach is that the model gets trained very quickly and the results are also obtained very fast as there is no training time spent like in training a neural model.

\begin{acks}
I express my gratitude to Bahadir Altun and Dr. Mucahid Kutlu (TOBB ETU team) for their implementation and code on the Github for the task.
\end{acks}

\bibliographystyle{ACM-Reference-Format}
\bibliography{sample-base}

\end{document}